\documentclass[
 reprint,
superscriptaddress,
 amsmath,amssymb,
 aps,
longbibliography
]{revtex4-2}

\usepackage{physics}

\usepackage{graphicx}
\usepackage{dcolumn}
\usepackage{bm}

\usepackage{xcolor}

\usepackage{mathtools}
\usepackage{gensymb}
\usepackage{lipsum}

\usepackage{caption}
\usepackage{subcaption}
\DeclareCaptionLabelSeparator{bar}{~\rule[-0.4ex]{0.2ex}{1em}~}
\DeclareCaptionLabelFormat{subfor}{\textbf{#2}}
\newcommand*\bfcaption[2]{\caption[#1]{\textbf{#1.}#2}}
\usepackage{xcolor}
\definecolor{UBcolor}{HTML}{007CC1}
\usepackage[colorlinks=true,pdfnewwindow=true,linkcolor=UBcolor,citecolor=UBcolor,urlcolor=UBcolor,breaklinks=true,linktocpage]{hyperref}
\usepackage[all]{hypcap}
\usepackage[nameinlink,capitalise]{cleveref}
\crefname{SI section}{SI Section}{SI Sections}
\Crefname{SI section}{SI Section}{SI Sections}

\begin{document}
\preprint{APS/123-QED}

\title{Accuracy Comes at a Cost: Optimal Localisation Against a Flow}
\author{Till Welker}
\email{t.a.welker@sms.ed.ac.uk}
\affiliation{School of Physics and Astronomy, University of Edinburgh}
\author{Patrick Pietzonka}
\email{p.pietzonka@ed.ac.uk}
\affiliation{School of Physics and Astronomy, University of Edinburgh}

\begin{abstract}
How much work does it cost for a propelled particle to stay localised near a stationary target, defying both thermal noise and a constant flow that would carry it away? We study the control of such a particle in finite time and find optimal protocols for time-dependent swim velocity and diffusivity, without feedback. Accuracy, quantified via the mean squared deviation from the target, and energetic cost turn out to be related by a trade-off, which complements the one between precision and cost known in stochastic thermodynamics. We show that accuracy better than a certain threshold requires active driving, which comes at a cost that increases with accuracy. The optimal protocols have discontinuous swim velocity and diffusivity, switching between a passive drift state with vanishing diffusivity and an active propulsion state. This study highlights how a time-dependent diffusivity enhances optimal control and sets benchmarks for cost and accuracy of artificial self-propelled particles navigating noisy environments.
\end{abstract}

\maketitle

\noindent\textit{Introduction -} A human swimmer in a river is fighting against the flow that drags her downstream. To stay in place, she needs to perform work to actively swim against the flow. The same is true for microscopic self-propelled particles in a flow. Being small generally helps to reduce the work required for propulsion, yet thermal fluctuations then pose an additional challenge. In this letter, we find the propulsion and diffusivity that optimise accuracy of localising a particle in a flow for any given work budget over a fixed time interval. Surprisingly, the best results are obtained for propulsion and diffusivity following non-trivial time protocols.

Our motivation to study this setup is two-fold. First, it presents a minimal way to analyse the role of accuracy in stochastic thermodynamics~\cite{seifert2012stochastic}. So far, the thermodynamic uncertainty relation establishes a trade-off between cost and \textit{precision}~\cite{barato2015thermodynamic,gingrich2016dissipation,hyeon2017physical,horowitz2017proof,horowitz2020thermodynamic}. There, precision is quantified via the mean square deviation of a stochastic variable from its own mean. Accuracy, on the other hand, relates to the mean square deviation from a desired target value. Being precise is not enough to be accurate, since a precisely located particle can still be far away from a desired target.

Trade-offs between cost and accuracy have been proposed for specific systems with feedback control~\cite{cocconi2025dissipation}. Here, as a minimal setting, we focus on control that does not require any measurements after the initialisation of the system. In the absence of flows, the problem of accurately localising a Brownian particle at a fixed position is trivial: Resting at a target at zero diffusivity yields perfect accuracy at zero cost. Hence, the simplest non-trivial question is how a particle can be close to a target that moves at a constant velocity. Or equivalently (after transformation to a co-moving frame~\cite{speck2008role}), how to localise at a stationary target while being subject to a constant flow. 

Beyond its theoretical value, the setup we consider can provide simple benchmarks for the cost of control of active particles in general.  To power their self-propulsion, active particles use fuel, which drives them out of equilibrium~\cite{zottl2016emergent,gompper20202020}. Individual active particles can carry cargo \cite{baraban2012transport,popescu2011pulling} and promise applications in targeted drug delivery \cite{patra2013intelligent,baylis2016halting,wang2018recent}, where accuracy is desirable. As a collective of active particles, active matter 
behaves in ways not possible at equilibrium; it can phase separate without attraction~\cite{cates2015motility,buttinoni2013dynamical,fily2012athermal} and flock directionally~\cite{vicsek1995novel,toner1998flocks}. By tuning activity and friction in time and space or imposing boundaries, one can control the direction of flocking~\cite{falk2021learning}, stabilise vortices~\cite{reinken2020organizing}, and accelerate self-assembly~\cite{schubert2025self}.  Such control of active matter harnesses its microscopically injected energy on a macroscopic scale and must come at a minimal cost, analogously to the results we obtain here for the control of individual active particles. 

As a systematic framework to optimise cost and accuracy in small systems, we use optimal control theory~\cite{blaber2023optimal,alvarado2025optimal,seifert2025stochastic}. There, control parameters are determined such that a predefined objective (travel time, work, accuracy, etc.) is optimised. The resulting optimal protocols can feature discontinuities~\cite{blaber2023optimal,blaber2021steps} or non-monotonicities~\cite{loos2024universal}. In active matter, optimal control theory has been used to extract work from active propulsion~\cite{garcia2024optimal}, translate an active droplet~\cite{shankar2022optimal}, move an active particle between two local minima~\cite{gupta2023efficient}, drive transitions between non-equilibrium steady states~\cite{baldovin2023control}, and transform a potential from an initial to a final shape~\cite{garcia2024optimal,davis2024active}.
Optimal control also helps devise strategies to reach a target in a flow. In noise-free systems, an elegant solution exists~\cite{liebchen2019optimal}, but it fails once noise is introduced. In response, strategies have been proposed that measure the environment and compensate the effect of noise~\cite{nava2018markovian,calascibetta2023optimal,piro2024energetic}, a task that can be learned by machine learning~\cite{schneider2019optimal,muinos2021reinforcement}.

We obtain a closed-form solution for the work-accuracy trade-off of a swimmer localising in a constant flow, which is arguably the simplest non-trivial accuracy optimisation problem in stochastic thermodynamics. The optimal protocols we find switch between passive drifting and active swimming, with discontinuous changes in velocity and diffusivity. As priority is shifted from high accuracy toward cost saving, a second-order phase transition from partially active to purely passive protocols occurs. Crucially, while previous approaches to control active particles treat their diffusivity as fixed, we show that time-dependent control of diffusivity (e.g., by varying shape or volume) significantly reduces cost, or allows for better accuracy at the same cost.

\begin{figure}
    \centering
    {\phantomsubcaption\label{Fig:schematic}}
    {\phantomsubcaption\label{Fig:pareto}}
    \includegraphics[scale=0.74]{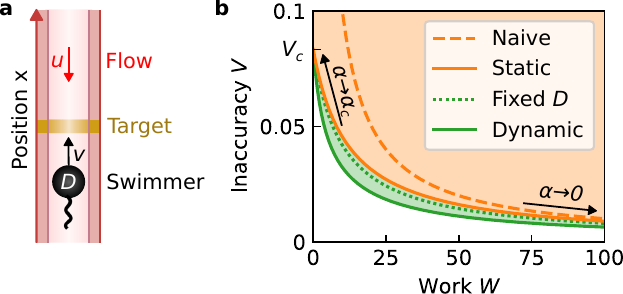}
    \bfcaption{System and trade-off}{  \subref*{Fig:schematic}, Swimmer with propulsion velocity $v(t)$ and diffusivity $D(t)$ in a flow of velocity $-u$ trying to stay close to a target at $x=0$. \subref*{Fig:pareto}, Optimal trade-off between work $W$ and inaccuracy $V$ represented by the Pareto front for static (orange line) and for dynamic (green line) protocols for $v(t)$ and $D(t)$. The trade-off parameter $\alpha$ controls the emphasis on reducing $W$ during optimisation. The shaded area represents the region accessible for suboptimal protocols. The naive protocol (orange dashed line) describes swimmers staying on average on the target. Protocols with dynamic velocity but fixed diffusivity lead to the Pareto front shown as the green dotted line.}
    \label{Fig:system_pareto}
\end{figure}

\bigskip
\noindent\textit{Model -} We study a propelled particle (``swimmer'') moving along a one-dimensional coordinate $x$ while being subject to an external flow with velocity $-u$, as shown in~\cref{Fig:schematic}. The swimmer's propulsion force $f(t)$ and mobility $\mu(t)$ can be adjusted by a (generally) time-dependent protocol, running from time $t=0$ to the total time $\tau$. The mobility is the inverse of the friction coefficient, which is determined by geometric features of the swimmer. A time-dependent mobility can be realised by varying size, shape or orientation, or by binding to and unbinding from other objects advected in the flow. The swimmer's velocity relative to the flow is then $v(t)=\mu(t)f(t)$  and its diffusivity is set by the Einstein relation to $D(t)=k_\mathrm{B}T\mu(t)$ at thermal energy $k_\mathrm{B}T$. On mesoscopic scales, inertia can be neglected, so that the swimmer's position in the lab frame follows the overdamped Langevin equation
\begin{align}
    \dot x(t) = v(t)-u+\sqrt{2D(t)} \xi(t) \,,
\end{align}
where $\xi(t)$ is a Gaussian white noise with zero mean and correlations $\langle \xi(t)\xi(t') \rangle = \delta(t-t')$. Starting from the initial position $x_0$, the probability distribution of the particle position is Gaussian with mean \mbox{$\bar{x}(t) = x_0 +\int^t_0 \dd t' \, (v(t)-u)$} and variance \mbox{$\sigma^2(t)= 2\int_0^t \dd t'\,D(t')$}. The system is dedimensionalised by setting protocol time, flow speed, and thermal energy to $\tau,u,k_BT=1$.

We aim for the swimmer to be close to a target at position $x=0$ over the whole time interval. The ensemble-averaged squared distance from the target at time $t$ is the second moment $\langle{x(t)^2}\rangle= \sigma^2(t) + \bar{x}^2(t)$ of the particle's probability distribution. Averaging this quantity over the protocol time leads to
\begin{align}\label{Eq:deviation}
    V &\equiv \int_0^1 \dd t \Big[\sigma^2(t) +\bar{x}^2(t)\Big] \nonumber\\
    &= \int_0^1 \dd t \Big[2D(t)\,(1-t) +   \bar{x}^2(t)\Big] \,,
\end{align}
which we call the ``inaccuracy''.  Small inaccuracy means good accuracy. The expression in the second line is obtained using integration by parts of the first term. It shows that diffusion impacts $V$ more strongly at the beginning of the time interval than at its end, because the effect of any fluctuation lasts for the remainder of the protocol time. 
The mean work required for swimming is~\cite{speck2008role},
\begin{align}\label{Eq:work}
    W \equiv \int_0^1\dd t\, f(t) v(t) = \int_0^1 \dd t\, \frac{(\dot {\bar{x}}(t)+1)^2}{D(t)} \,,
\end{align}
where $v(t)=\dot {\bar{x}}(t)+u$ (with our convention $u=1$) is the swim velocity.
Equations~\eqref{Eq:deviation} and~\eqref{Eq:work} already hint at a trade-off between work and inaccuracy. Work $W$ can be reduced by a slower swim velocity $v(t)$ or larger diffusivity $D(t)$. Yet, both lead to larger inaccuracy, either through an increased distance between the mean and the target, or through an increased variance.

We obtain the protocols that optimise the trade-off between inaccuracy $V$ and mean work $W$, by minimising the utility functional~\cite{ngatchou2005pareto}
\begin{align} \label{Eq:utility_functional}
    F_\alpha[D,\bar{x}] = V[D,\bar{x}] + \alpha W[D,\bar{x}]
\end{align}
over $D$ and $\bar{x}$. Here, we have made the functional dependence of $V$ and $W$ on $D(t)$ and $\bar{x}(t)$ explicit. The mean position in the lab frame $\bar{x}(t)$ encapsulates both the controlled force $f(t)=(\dot{\bar{x}}(t)+1)/D(t)$ and the initial condition $x_0=\bar x(0)$. The trade-off parameter $\alpha>0$ prioritises accuracy (small $V$) when it is small and fuel saving (small $W$) when it is large. By varying $\alpha$ we obtain the Pareto front~\cite{ngatchou2005pareto,seoane2016multiobjective} for the trade-off between $V$ and $W$ shown in \cref{Fig:pareto}. From it, one can read off the best accuracy achievable for a given fuel budget.

\bigskip
\noindent\textit{Static Protocol -} We start by considering protocols with fixed swim velocity $v$ and diffusivity $D$, which are straightforward to implement in experiments and provide a reference for the time-dependent protocols considered below. Inserting the diffusivity $D$ and the mean $\bar{x}(t)=x_0+(v-1)t$ into the utility functional \eqref{Eq:utility_functional} yields the function
\begin{align} \label{Eq:utility_function}
    F_\alpha(D,v,x_0) = D + \frac{(v-1)^2}{3} + x_0 (v-1)+x_0^2+\alpha\frac{v^2}D\,.
\end{align}
By solving $\partial F_\alpha/\partial D=0$ and $\partial F_\alpha/\partial {x_0}=0$, we obtain the optimal diffusivity $D^*=\sqrt\alpha |v|$ and initial condition $x_0^* =(1-v)/2$ that minimise $F_\alpha$ for given~$v$ . Note that the swimmer is optimally initialised such that its time-averaged mean is zero. Inserting $D^*$ and $x_0^*$ in \cref{Eq:utility_function} gives the utility function \mbox{$F_\alpha(v)=2\sqrt{\alpha}|v| +(v-1)^2/12$}, which has a kink at $v=0$. Depending on $\alpha$, this function can be minimal either at the non-zero swim velocity where $\partial F_\alpha/\partial v$ vanishes, or at the kink. We obtain the optimal velocity $v^*=1-12\sqrt{\alpha}$ for $\alpha\leq \alpha_\mathrm{c} =1/144$ and $v^*=0$ otherwise. 

\begin{figure}
    \centering
    {\phantomsubcaption\label{Fig:constant_velocity}}
    {\phantomsubcaption\label{Fig:constant_diffusivity}}
    {\phantomsubcaption\label{Fig:constant_schematic}}
    \includegraphics[scale=0.74]{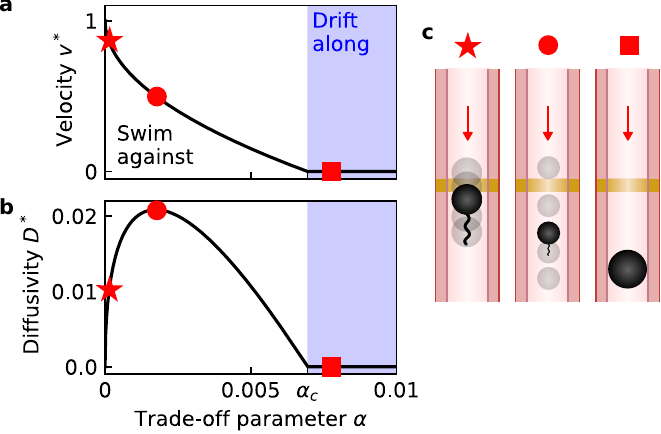}
    \bfcaption{Static Protocol}{ \subref*{Fig:constant_velocity}, Optimal swim velocity and, \subref*{Fig:constant_diffusivity}, optimal diffusivity as a function of the trade-off parameter $\alpha$. \subref*{Fig:constant_schematic}, schematic illustration of expensive $\bigstar$, intermediate \raisebox{-0.4ex}{\LARGE$\bullet$} and free $\blacksquare$ protocols. The solid black particles indicate the mean at the end of the protocol, while the transparent black particles illustrate the diffusive spread. The different sizes of particles and flagella represent the changed mobility (diffusivity) and swim velocity, respectively.}
    \label{Fig:constant}
\end{figure}

The optimal swim velocity and diffusivity depend characteristically on the trade-off parameter $\alpha$, as shown in \cref{Fig:constant}a,b. 
With increasing $\alpha$, the optimal swim velocity monotonically decreases from the flow velocity $u=1$ to $0$, while the optimal diffusivity first increases from $0$ to about $0.02$, before decreasing again to 0. This behaviour can be understood by considering the three protocols sketched in \cref{Fig:constant_schematic}. For expensive protocols, the diffusivity (and hence mobility) can be small while actively swimming at a velocity close to $u$ against the flow. For protocols of intermediate cost, one chooses smaller velocity and larger diffusivity. Beyond $\alpha_\mathrm{c}$, the optimal velocity and diffusivity are zero. This protocol is free of any cost, it avoids thermal fluctuations, but results in a large average deviation as the swimmer passively drifts past the target.
The transition at $\alpha_\mathrm{c}$ from active swimming against the flow to passive drifting along is akin to a second-order phase transition with a jump in $\partial_\alpha v^*$. This is a common feature of multiobjective optimisation problems~\cite{seoane2016multiobjective,sunil2024minimizing}.

By inserting the optimal static protocol into \cref{Eq:deviation,Eq:work}, we obtain the inaccuracy $V^* = \sqrt{\alpha}$ and work $W^* = 1/\sqrt{\alpha} -12$ for $\alpha \leq \alpha_\mathrm{c}$.
This can be rearranged to
\begin{align} \label{Eq:pareto_const}
    V^*(W)= \frac{1}{W+12}
\end{align}
giving the minimal inaccuracy $V^*$ for a given fuel budget $W$ using a static protocol. This optimal trade-off is shown in \cref{Fig:pareto}. Note that the inaccuracy \mbox{$V_\mathrm{c}=V^*(W=0)=1/12$} is finite even if no work is performed, as the particle drifts along with the flow for a finite time. This protocol is much better than the naive protocol of initialising the swimmer at the target $x_0=0$, swimming at the velocity of the flow $v=1$, and tuning the cost through the diffusivity $D$. For this naive protocol one gets $V =D$ and $W =1/D$, such that inaccuracies  $V=1/W$ diverge as the work goes to zero.

Let us examine the trade-off using dimensionful quantities for the accuracy $\tilde V$, the flow velocity $\tilde u$, and the protocol time $\tilde \tau$. For fixed $\tilde V$, the dimensionless accuracy $V=\tilde V/(\tilde{u}\tilde{\tau})^2$ decreases for increasing flow velocity $\tilde u$ and protocol time $\tilde \tau$. Achieving the same dimensionful accuracy in a faster flow or for longer times requires smaller $V$ and hence more work (measured in $k_BT$). As an illustrative example we plug in a flow velocity $\tilde u = 10\,\mathrm{\mu m/s}$ (typical swimming speed of bacteria) and a time $\tilde\tau = 1\,\mathrm{s}$, resulting in a characteristic length $\tilde l =\tilde u \tilde \tau = 10\,\mathrm{\mu m}$. To reach an inaccuracy less than $\tilde V_\mathrm{c} = \tilde l^2 V_\mathrm{c}\simeq (2.9 \,\mathrm{\mu m})^2$, active swimming is necessary. To localise the swimmer to $\tilde V=1\,\mathrm{\mu m}^2$,
\cref{Eq:pareto_const} dictates a minimal work of $\tilde W = 88\,k_BT$. The optimal swimmer has velocity $\tilde v = 8.8 \, \mathrm{\mu m/s}$ and diffusivity $\tilde D = 0.88 \, \mathrm{\mu m^2/s}$. For a spherical swimmer in a liquid with viscosity $\tilde\eta=1\,\mathrm{mPa\,s}$ (similar to water) the corresponding radius given by the Stokes–Einstein–Sutherland equation is $\tilde r = k_B T /(6\pi \tilde\eta \tilde D)\simeq 0.25\,\mathrm{\mu m}$. Swimmers of such size can be built, as demonstrated by porous drug carrying active Janus particles reaching sizes below $100\,\mathrm{nm}$~\cite{ma2015catalytic}.

\bigskip
\noindent\textit{Dynamic Protocols -}  To further improve the work–accuracy trade-off, one can implement time-dependent protocols for the control parameters of the model. For instance, one can consider a setup where the propulsion force and hence the swim velocity $v(t)$ are controlled in a time-dependent way, while the shape of the swimmer and hence its diffusivity are time-independent. The corresponding Pareto front in \cref{Fig:pareto} (labeled ``fixed $D$'') shows that such a dynamic protocol does improve the accuracy compared to the static protocols for any given work budget. However, this improvement is rather small. The calculations leading to this Pareto front are shown in Appendix~\ref{sec:fixed_velocity}, they follow similar steps as in the more general setup we here focus on instead. We consider simultaneous time-dependent control of  both the velocity $v(t)$ and the diffusivity $D(t)$ of a swimmer. We now demonstrate that this kind of control leads to a significant advantage in the trade-off.

The goal of the optimisation is again to minimise the utility functional $F_\alpha$ in \cref{Eq:utility_functional}, where we now also admit time-dependent protocols $D(t)$, $\bar{x}(t)$. Recall the relation of the latter to the swim velocity $v(t)= \dot{\bar{x}}(t)+1$.
Straightforward optimisation via $\delta F_\alpha/\delta D = 0$ gives the diffusivity 
\begin{align}\label{Eq:optimal_D}
    D^*(t)=\sqrt{\frac{\alpha}{2(1-t)}}|\dot {\bar{x}}(t)+1| \,.
\end{align}
The scaling $D^*\propto \sqrt{\alpha}|v(t)|$ resembles the static case: a large diffusivity mitigates cost, as required for fast swimming, and if $\alpha$ puts a strong priority on reducing cost. The additional factor $1/\sqrt{2(1-t)}$ reduces fluctuations at the start of the protocol, and increases them towards the end. This is beneficial, because early fluctuations worsen the overall accuracy for the remainder of the protocol time. 

From now on, we use this optimal diffusivity $D^*(t)$ and search for the optimal protocol for the mean position $\bar{x}^*(t)$. Inserting $D^*(t)$ of \cref{Eq:optimal_D} in \cref{Eq:utility_functional} yields the functional depending only on $\bar{x}(t)$ 
\begin{align}\label{Eq:utility_functional_u}
        &F_\alpha[\bar{x}] =\int_0^1 \dd t\, L(\bar{x}(t),\dot{\bar{x}}(t),t)
\end{align}
with the Lagrangian 
\begin{align}\label{Eq:Lagrangian}
    L(\bar{x},\dot{\bar{x}},t) = 2\sqrt{2\alpha(1-t)}|\dot {\bar{x}}+1|+ \bar{x}^2.
\end{align}
The corresponding Euler-Lagrange equation is
\begin{align*}
    0 &= \frac{\delta F_\alpha}{\delta \bar x} = \frac{\partial L}{\partial \bar x} - \frac{\dd}{\dd t}\frac{\partial L}{\partial \dot{\bar x}} \\
    &= 2 \bar x - \frac{\dd}{\dd t} 2 \sqrt{2\alpha(1-t)} \,\mathrm{sign}(\dot{\bar x} +1).
\end{align*}
Requiring $\dot{\bar x} +1>0$ yields as candidate for the optimal protocol
\begin{align} \label{Eq:Mean_Swim}
    \bar{x}_\mathrm{s}(t) = -\sqrt{\frac{\alpha}{2(1-t)}} ~~~\text{for} ~~ t< t_\mathrm{max} = 1-\frac{\alpha^{1/3}}{2} \,,
\end{align}
which is consistent with the requirement only up to the time $t_\mathrm{max}$.
Here, the swimmer actively swims at velocity \mbox{$v_\mathrm{s}(t)=\dot {\bar{x}}_\mathrm{s}(t)+1 =1-\sqrt{2\alpha}(1-t)^{-3/2}/4>0$} against the flow. For further candidates for optimal protocols, note that the Lagrangian has a kink at $\dot {\bar{x}}+1=0$, which can be a minimum for a given time $t$. Consequently, protocols $\bar{x}_\mathrm{d}(t)=c - t$ with arbitrary offset $c$, where the swimmer drifts passively with the flow, can also be optimal in parts of the protocol. 

\begin{figure}
    \centering
    {\phantomsubcaption\label{Fig:optimal_protocol}}
    {\phantomsubcaption\label{Fig:glue_times}}
    \includegraphics[scale=0.695]{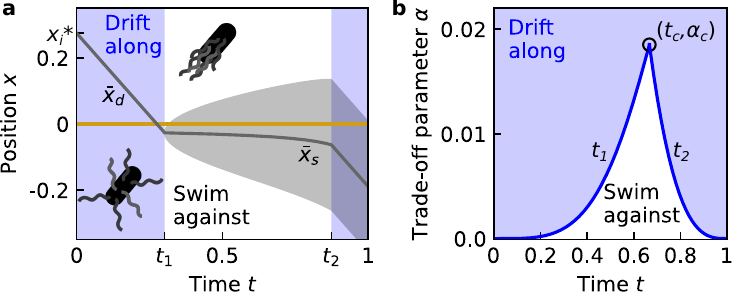}
    \bfcaption{Dynamic protocol}{ \subref*{Fig:optimal_protocol}, Optimal protocol for trade-off parameter $\alpha = 0.001$. The grey line is the optimal mean $\bar{x}^*(t)$, the grey area is the standard deviation caused by the optimal diffusivity $D^*(t)$, and the golden line is the target. The drifting phase with zero mobility is coloured in blue and illustrated by a bacterium with spread flagella. The swimming phase is coloured in white and illustrated by a bacterium with bundled flagella. \subref*{Fig:glue_times}, Optimal switching times $t_1$ and $t_2$ over trade-off parameters $\alpha$. When $\alpha$ is increased, the cost is reduced by shortening the swimming period. At the critical value $\alpha_\mathrm{c}$, the start and end of the swim period meet at $t_\mathrm{c} = 2/3$. Beyond $\alpha_\mathrm{c}$, it is optimal to always drift along.
    }
\end{figure}

The truly optimal protocol $\bar{x}^*(t)$ for each trade-off parameter $\alpha$ must be pieced together from the swim protocol $\bar{x}_\mathrm{s}(t)$ and the appropriately shifted drift protocols $\bar{x}_\mathrm{d}(t)$ , as shown in \cref{Fig:optimal_protocol}. Provided that the optimal protocol $\bar{x}(t)$ has no discontinuous jumps, the most general ansatz for it is: the swimmer starts drifting until it reaches $\bar{x}_\mathrm{s}(t)$ at time $t_1$, then it swims until the time $t_2$, and afterwards continues drifting until the end of the protocol (note that it can never reach $\bar{x}_\mathrm{s}(t)$ again, because $v_\mathrm{s}(t)>v_\mathrm{d}=0$ for $t<t_\mathrm{max}$). The initial position $x_\mathrm{i}$ and final mean position $\bar x_\mathrm{f}$ for this ansatz are determined by $t_1$ and $t_2$ as $x_\mathrm{i} =t_1+\bar{x}_\mathrm{s}(t_1)$, $\bar x_\mathrm{f}=1-t_2+\bar{x}_\mathrm{s}(t_2)$. The assumption that there are no jumps during the protocol is validated by mapping the Lagrangian optimisation to a Hamiltonian problem and studying its dynamics in Appendix~\ref{Sec:Hamiltonian}.   

Inserting the drift-swim-drift ansatz into the utility functional in \cref{Eq:utility_functional_u} yields a function $F(t_1,t_2)$. In Appendix~\ref{Sec:Construct_Solution} we show that optimising this function gives an explicit expression $t_2=1-(2\alpha)^{1/3}$ for the second switch time and an implicit expression \mbox{$\alpha=(t_1^4-t_1^5)/(8-8t_1+2t_1^2)$} for the first switch time. Hence, the optimal dynamic protocol can be expressed explicitly in terms of $t_1$, instead of $\alpha$. The resulting optimal switching times are shown in \cref{Fig:glue_times}. When the trade-off parameter $\alpha$ is increased, the cost is reduced by shortening the swim duration. At a critical value $\alpha_\mathrm{c}=1/54$, the start and end points of the swimming phase meet at $t_\mathrm{c} =2/3$. For $\alpha>\alpha_\mathrm{c}$ it is best to always drift along. This transition corresponds to a second-order phase transition, similar to the result for static protocols, but at a larger value of $\alpha_\mathrm{c}$.

We can intuitively understand the properties that characterise an optimal protocol, such as the one shown in \cref{Fig:optimal_protocol}. The swimmer starts upstream and drifts along, with a vanishing mobility (and hence vanishing diffusivity). This saves cost early in the protocol, and avoids fluctuations that would contribute to the inaccuracy over its remaining duration. Soon after the swimmer passes the target, it actively swims against the flow to stay close to it. Swimming at finite cost requires non-zero mobility, and consequently, diffusive fluctuations cannot be avoided. Towards the end of the protocol, the swimmer drifts away with zero mobility, which again avoids cost, while the increasing displacement only affects a short time period.

Inserting the optimal protocols into the functionals for inaccuracy $V$ (\cref{Eq:deviation}) and work $W$ (\cref{Eq:work}) yields the Pareto front shown in \cref{Fig:pareto}. These dynamical protocols perform significantly better than the static protocols. Revisiting the example presented above: to reach an inaccuracy of $V=0.01$, the static protocol requires a work of $88\, k_B T$ while the dynamic protocol only requires $57.3\, k_BT$.

\vspace{1.5mm}
\noindent\textit{Conclusion -} In this letter, we have revealed a trade-off between work and accuracy for a microswimmer that aims to localise near a target while being subject to a constant flow. The optimal time-dependent strategies switch between a passive drift state and an active swim state, with discontinuous jumps in swim velocity and diffusivity during the protocol.

Our findings identify swim velocity and diffusivity as powerful control handles for optimal navigation strategies, which so far had focused on the control of swim direction. Even if velocity and diffusivity are fixed during the experiment, our optimal static protocols guide experimentalists when tailoring colloid size and swim speed to the environment's flow speed and protocol time.

The proof of principle established here opens the door to new questions.
Extending the model to two and three dimensions, accounting for non-uniform or time-dependent flows, and incorporating thermodynamically consistent feedback control are natural next steps. Furthermore, experimental realisations of tunable diffusivity, such as shape or volume changes, or (un)binding from passive objects advected in the flow, should be explored. For most cases, there will be practical limitations on the range of control parameters. Addressing these limitations, as well as uncertainties in the initial conditions and the environment, can pave the way toward real-life applications. Even if the idealised full control of diffusivity and swim speed cannot be achieved in those applications, our results provide theoretical limits on cost and accuracy that serve as ultimate benchmarks.

\onecolumngrid
We thank Martin Evans for discussions and comments on the manuscript.
\twocolumngrid

\bibliography{Bibliography}

\newpage
\phantom{.}
\newpage
\appendix

\section{Dynamic velocity and fixed diffusivity}\label{sec:fixed_velocity}
\begin{figure}
    \centering
    {\phantomsubcaption\label{Fig:fixedD_optimal_D}}
    {\phantomsubcaption\label{Fig:fixedD_trajectory}}
    \includegraphics[scale=0.7]{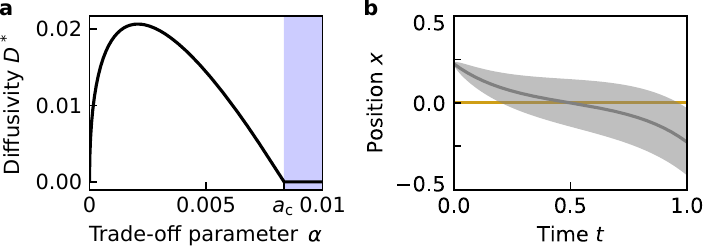}
    \bfcaption{Dynamic velocity and fixed diffusivity}{  \subref*{Fig:fixedD_optimal_D}, Optimal diffusivity $D^*$ as a function of the trade-off parameter $\alpha$. \subref*{Fig:fixedD_trajectory}, Optimal protocol with dynamic velocity $v(t)$ but constant diffusivity $D$ for $\alpha =0.001$. As in \cref{Fig:optimal_protocol}, the grey line is the optimal mean, the grey area is the standard deviation caused by the optimal diffusivity, and the golden line is the target.}\label{Fig:fixed_D}
\end{figure}

In this section, we discuss the scenario in which the velocity $v(t)$ can be varied, while the diffusivity $D$ is fixed throughout the protocol.
In this case, the inaccuracy $V$ and the work $W$ given in \cref{Eq:deviation,Eq:work} simplify to:
\begin{align}\label{eq:inaccuracy_work_fixedD}
    V = D + \int_0^1 \dd t\,    \bar{x}^2(t) \,,~
    W = \frac{1}D \int_0^1 \dd t\, (\dot {\bar{x}}(t)+1)^2 \,.
\end{align}
The utility functional $F = V + \alpha W$ in \cref{Eq:utility_functional} can be optimised with respect to the mean position $\bar x$ by solving the Euler-Lagrange equation
\begin{align*}
    0 = \frac{\delta F_\alpha}{\delta \bar x} = 2 \bar x(t) - 2 \frac{\alpha}{D} \ddot{\bar x }(t)\,.
\end{align*}
The ansatz for the solution reads
\begin{align}\label{Eq:fixed_diffusivity}
    \bar{x}(t) = c_1 e^{-\lambda t} +c_2 e^{\lambda t}\,,
\end{align}
with $\lambda = \sqrt{D/\alpha}$. Inserting  this in the inaccuracy and work in \cref{eq:inaccuracy_work_fixedD} results in a utility function $F_\alpha(c_1,c_2,D)=V+\alpha W$. The utility function is optimised with respect to $c_i$ by solving $\partial F_\alpha/\partial c_i = 0$, which yields $c_1 = e^{\lambda}/[\lambda(1+e^\lambda)]$ and $c_2 = -1/[\lambda(1+e^\lambda)]$. Inserting the optimal $c_i$ into \cref{Eq:fixed_diffusivity} results in the optimal mean trajectory,
\begin{align*}
    \bar{x}^*(t) &= \frac{1}{\lambda(1+e^\lambda)} \Big(e^{\lambda (1-t)} -  e^{\lambda t} \Big) \,.
\end{align*}
To obtain the optimal diffusivity, we insert $c_i$ in the utility function, which yields
\begin{align*}
    F_\alpha(D) =  D + \frac{\alpha}{D}- \frac{2\alpha^{3/2}}{D^{3/2}}\frac{e^{\sqrt{D/\alpha}} -1}{e^{ \sqrt{D/\alpha}}+1}\,.
\end{align*}
We numerically minimise the function for fixed $\alpha$ to obtain the optimal $D$ shown in \cref{Fig:fixedD_optimal_D}. As for the static protocol shown in \cref{Fig:constant}, the diffusivity non-monotonically depends on the trade-off parameter $\alpha$. At a critical value $\alpha_c$, there is a second-order phase transition with a discontinuity in $\partial D^*/\partial\alpha$. Beyond $\alpha_c$, it is best to drift by setting swim velocity and diffusivity to zero. As $\alpha$ approaches the critical value and the diffusivity approaches $0$, we can calculate $\lim_{D\to 0} \partial F_{\alpha}/\partial D = 1-1/(120\alpha)$ using L'Hôpital's rule, which yields $\alpha_\mathrm{c} = 1/120$. Note that the critical value $\alpha_c$ is larger than for the static protocol but smaller than for the fully dynamic protocol.

Inserting the optimal parameters into $W$ and $V$ yields the work-accuracy trade-off shown in \cref{Fig:pareto}. The protocol with dynamic velocity but static diffusivity performs slightly better than the one in which both parameters are static, but significantly worse than the one in which both are dynamic. This highlights the benefits of tunable diffusivity.

The optimal protocol for $\alpha = 0.001$ is shown in \cref{Fig:fixedD_trajectory}. Remarkably, $\bar{x}^*(t)$ is symmetric under simultaneous time and parity inversion. To understand why, we adapt the symmetry argument in Ref.~\cite{loos2024universal}. Consider a mean trajectory $q(t)$ and its time and parity reversed image $\tilde q(t)=-q(1-t)$. Both work and inaccuracy in Eq.~\eqref{eq:inaccuracy_work_fixedD} are invariant under this transformation. Because the functional is quadratic in $q$ and $\dot {q} $, we expect a unique extremum with $q^*(t)=\tilde q^*(t)$. The invariance of the utility under the inversion breaks for time-dependent diffusivity, as becomes apparent from the Lagrangian in \cref{Eq:Lagrangian}. As a consequence, the optimal protocols for dynamic $D(t)$ are no longer symmetric under time and parity inversion as shown in~\cref{Fig:optimal_protocol}.

\section{Hamiltonian formulation} \label{Sec:Hamiltonian}
\begin{figure}
    \centering
    {\phantomsubcaption\label{Fig:hamilton_potential}}
    {\phantomsubcaption\label{Fig:hamilton_trajectory}}
    \includegraphics[scale=0.7]{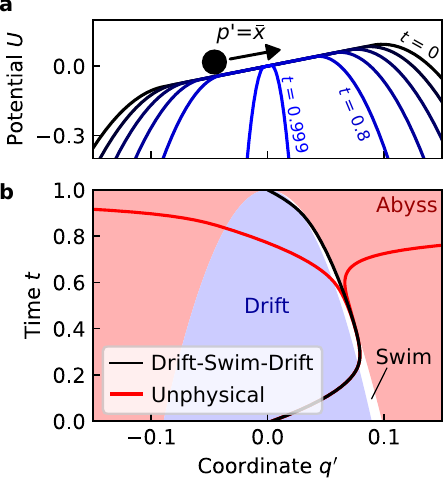}
    \bfcaption{Hamiltonian Description}{ \subref*{Fig:hamilton_potential}, The Lagrangian optimisation problem can be mapped to a ball with momentum $p' = \bar x$ in a time-dependent potential. The linear part of the potential corresponds to the drift protocol, and the non-linear region between the linear region and the location of the maximal potential corresponds to the swim protocol. The abyss beyond drift and swim protocols results in divergences for $\epsilon \to 0$. \subref*{Fig:hamilton_trajectory}, Swim, drift, and abyss region. The black line is the solution to the Hamiltonian problem with $p'(0)=x_0$ and $q'(1)=0$ (for which $q'(0)\approx0.00357$). The solution is sensitive to initial conditions, as shown by the red curves, generated with $q'(0)=0.00355,0.00360$. Here we chose $\alpha = 0.001$ and $\epsilon =0.1$ and integrate the dynamics using the Julia package \textit{DifferentialEquations}~\cite{rackauckas2017differentialequations}.}
\end{figure}

In this section, we map the Lagrangian optimisation problem [minimise $F_\alpha = \int_0^1 dt \, L(q,\dot q,t)$] stated in \cref{Eq:utility_functional_u} to a Hamiltonian problem [solve the equations of motion of the Hamiltonian $H(q,p,t)$]. We start from the Lagrangian
\begin{align*}
    L(q,\dot q , t) =2\sqrt{2\alpha(1-t)}\left(|\dot q+1|+\frac\epsilon2(\dot q +1)^2\right)+q^2,
\end{align*}
writing the mean position as $\bar x = q$. There, we have added a regularisation term $O(\epsilon)$, which is required for a mapping from $\dot q$ to conjugated momenta $p$ covering all real numbers. For $\epsilon \to 0$, the original Lagrangian of \cref{Eq:Lagrangian} is recovered. The conjugated momentum is $p = \partial L/\partial \dot q = b(t)[\mathrm{sign}(\dot q +1)+\epsilon (\dot q +1)]$ with $b(t)=2\sqrt{2\alpha(1-t)}$. The Hamiltonian then follows as the Legendre transform
\begin{align*}
    H(q,p,t) = p\dot q - L = \frac{(|p| -b(t))^2}{2\epsilon b(t)} \theta(|p| -b(t)) -p -q^2 \,.
\end{align*}
When exchanging momentum and coordinate ($p\to q'$, $q\to p'$, and $H \to -H'$), we obtain a Hamiltonian with the mechanical interpretation of a point-like ball in a time-dependent potential,
\begin{align*}
    H'(q',p',t) = -\frac{(|q'| -b(t))^2}{2\epsilon b(t)} \theta(|q'| -b(t)) +q' +p'^2\,.
\end{align*}
The first terms are a time-dependent potential $U(q',t)$, sketched in \cref{Fig:hamilton_potential}, while the last term corresponds to a kinetic energy $p'^2/2m$ with mass $m=1/2$. After the mapping, the mean position of the original problem corresponds to the momentum $p'=\bar x$. Therefore, the initial momentum is set by the initial position $p'(0)=x_0$. But for a unique solution, we require an additional constraint, which we find by studying the potential $U(q',t)$.

In the region between $-b(t)$ and $b(t)$, the potential has a constant slope, such that \mbox{$\dot{\bar x} = \dot p' = -\partial_{q'}U = -1$}. This region corresponds to the drift protocol, where the swimmer drifts along with the flow. Beyond this region, the potential additionally decreases quadratically with a prefactor of order $O(1/\epsilon)$. This gives a monotonic potential for $q'<0$, while for $q'>0$ there is a maximum at $q_\mathrm{max} = b(t)(1+\epsilon)$. Between the connection point $b(t)$ and the maximum $q_\mathrm{max}$ there is a region of a width of order $O(\epsilon)$, in which the acceleration $\dot{\bar x} = \dot p' = -\partial_{q'}U$ is between $-1$ and $0$. This region corresponds to the swim protocol where the particle actively swims against the flow. The regions of drift and swim protocol are shown in \cref{Fig:hamilton_trajectory}.
Beyond $-b(t)$ and $q_\mathrm{max}(t)$ the particle is accelerated outward with acceleration of order $O(1/\epsilon)$. Because $-b(t)$ and $q_\mathrm{max}(t)$ constantly move toward $0$, a particle crossing the boundary outwards can never come back, and in the limit $\epsilon \to 0$, the acceleration $O(1/\epsilon)$ results in unphysical divergences. Therefore, a physical solution is confined to the interval $q'\in [-b(t),q_\mathrm{max}(t)]$.  For $t=1$, we have $-b(1)=q_\mathrm{max}(1) = 0$, and consequently $q'(1)=0$. Now we have a fully defined Hamiltonian problem with fixed initial momentum $p'(0)=x_0$ and final coordinate $q'(1)=0$. The solution for $\alpha = 0.001$, $\epsilon =0.1$ is shown in \cref{Fig:hamilton_trajectory}. Such a solution always exists because (1) there exist trajectories over-and undershooting $q'(1)=0$ and (2) the final position depends continuously on the initial condition for our system.

Solving this problem for $\epsilon \to 0$ gives the optimal mean particle trajectory $\bar{x}(t)=\lim_{\epsilon \to 0}p'(t)$ of the original problem. Note that in the drift and swim regions the acceleration is finite, while the outer region results in unphysical divergences for $\epsilon \to 0$. In contrast, a solution with a finite jump during the protocol would require an instance of diverging acceleration followed by finite acceleration of the ball. This is not possible in our system. Therefore, the optimal solution cannot have jumps in $\bar{x}(t)$ during the protocol.

The Hamiltonian formulation also tells us that there can be at most one swim phase. For the ball to leave the swim region towards the drift region, its velocity needs to be more negative than the velocity of the connection point $\dot b(t)$. In the drift region, the ball feels a constant (negative) acceleration, while the acceleration of the connection point  $\ddot b(t)$ gets monotonically more negative (and hence also the acceleration of the  maximum $\ddot q_\mathrm{max}(t)$). 
If the connection point reaches the ball again, it therefore must have a more negative velocity and more negative acceleration than the ball. And when the ball re-enters the swim region, its acceleration becomes even less negative and the maximum $q_\mathrm{max}(t)$ surpasses the ball in a time of order $O(\epsilon)$. As soon as the ball surpasses the maximum, it can not come back, and the result diverges for $\epsilon \to 0$. Consequently, a physical solution can have at most one swim phase.

\section{Constructing the optimal protocol} \label{Sec:Construct_Solution}
We insert the drift-swim-drift ansatz in the utility functional \eqref{Eq:utility_functional_u}, using the  drift protocols $\bar x_{d,1}(t)= t_1+\bar{x}_\mathrm{s}(t_1) -t$, $\bar{x}_{d,2}(t) = t_2+\bar{x}_\mathrm{s}(t_2) -t$,  and the swim protocol $\bar{x}_\mathrm{s}$ given in \cref{Eq:Mean_Swim}. Integration then gives
\begin{align}\label{Eq:functional_ansatz_optimal_initial}
    &F_\alpha(t_1,t_2)= F_1(t_1) + F_2(t_2) ~~\text{with}~~t_1\leq t_2\nonumber\\
    &F_1(t_1) = \frac{\alpha}{2} \frac{t_1}{1-t_1} - \sqrt\frac{\alpha}{2(1-t_1)} t_1^2 +\frac{t_1^3}{3}  \nonumber\\
    &\hspace{2cm} + \frac{4\sqrt{2\alpha}}{3} (1-t_1)^{3/2}-\frac{\alpha}{2}\ln(1-t_1) \nonumber\\
    &F_2(t_2) =  \frac13 (1-t_2)^3 -\frac{4\sqrt{2\alpha}}{3} (1-t_2)^{3/2} + \frac{\alpha}{2}\ln(1-t_2)\,.
\end{align}
The switching times $t_1$ and $t_2$ can be optimised separately, as long as $t_1\leq t_2$.
To obtain the optimal $t_2$, we solve $\partial_{t_2}F_\alpha=0$ resulting in two real solutions \mbox{$t_2^+ = t_\mathrm{max}$} and \mbox{$t_2^* = 1 - (2\alpha)^{1/3}$}. Inserting in \cref{Eq:functional_ansatz_optimal_initial} reveals $F_2(t_2^*)=a(\ln(2a)-3)/6<F_2(t_2^+)$, such that $t_2^*$ is optimal. The optimal $t_1$ is the solution to
\begin{align*}
    0=\partial_{t_1}F_\alpha = \partial_{t_1}F _1&=\frac{2-t_1}{2(1-t_1)^2} \alpha + t_1^2 - \Big(2\sqrt{2(1-t_1) }\\
    &+\sqrt{\frac{2}{1-t_1}}t_1 + (2(1-t_1))^{-3/2}\,t_1^2\Big)\sqrt{\alpha} \,.
\end{align*}
Note that $\partial_{t_1}F_\alpha$ is a second-order polynomial in $\sqrt{\alpha}$. The implicit solution $\alpha^* = \frac{t_1^4-t_1^5}{8-8t_1+2t_1^2}$ is the only one satisfying $t_1\leq t_2^*$.

\end{document}